\documentstyle[12pt,aaspp4]{article}

\def\insertplot#1#2#3#4#5#6#7{
\vskip 10pt\nobreak\hbox to \hsize{\hss\dimen0=#3in\hbox to #6\dimen0{%
\dimen0=#2in\vbox to #6\dimen0{\vss
\special{ps: plotfile #1}
\special{ps::[end]
  PGPLOT restore
}
}\hss}\hss}\vskip 10pt}
 
\begin{document}
\title{The Transitional PMS Object DI Tauri: Evidence for \\ 
a Sub-stellar Companion and Rapid Disk Evolution}
\author{Michael R. Meyer, S.V.W. Beckwith, T.M. Herbst, and M. Robberto\altaffilmark{1}}
\affil{Max--Planck--Institut f\"ur Astronomie, K\"onigstuhl 17, Heidelberg, Germany}

\altaffiltext{1}{Also at Osservatorio Astronomico di Torino, Italy.}

\begin {abstract} 

We report mid--IR observations of two young stars 
found in the Taurus dark cloud spatially resolving for the first time 
their 10 $\mu$m emission. The weak--emission T Tauri star 
DI Tau, tentatively identified by Skrutskie {\it et al.} (1990) 
on the basis of 12 $\mu$m IRAS data as an object in the process of 
dissipating its circumstellar disk, 
is found to have no infrared excess at a wavelength of 10 $\mu$m.
The nearby classical T Tauri star 
DH Tau exhibits excess emission at 10 $\mu$m consistent with
predictions based on circumstellar disk models.  While both 
objects appear to have the same stellar mass, age, and 
rotation rate, they differ in two fundamental respects: 
DH Tau is a single star with an active accretion disk 
and DI Tau is a binary system lacking such a disk.  The companion to 
DI Tau has a very low luminosity and is located 
at a projected distance of $\sim 20$ A.U. from 
the primary.  Assuming the system to be co--eval, we derive a 
mass below the hydrogen burning limit for the companion.
We speculate that the formation of a sub--stellar mass companion
has led to the rapid evolution of the circumstellar disk 
that may have surrounded DI Tau.  

\end{abstract}


\section {Introduction}

It is generally accepted that circumstellar
disks are a common by--product of the star formation
process (Beckwith and Sargent, 1996).  Estimates of the ubiquity of accretion 
disks around young stars ranges from $\sim$ 70 \% in the youngest
clusters (Carpenter {\it et al.}, 1997) to $\sim$ 50 \% in the Taurus dark cloud 
(Kenyon and Hartmann, 1995).  As young stars age, evidence of active
disk accretion diminishes (Hartigan, Edwards, and Ghandour, 1995). 
However, the process by which these disks dissipate 
remains a mystery. One possibility is that such disks give rise to the formation of 
planetary systems, though this has yet to be demonstrated. Even 
the influence of companion stars on the evolution of
circumstellar disks is unclear. 
Combining observations of 2.2 $\mu$m excess emission (originating
within a few stellar radii) with ground--based 10 $\mu$m and 
IRAS observations of mid--IR excesses (originating in the terrestrial planet region
from 0.1-2.0 A.U.) 
of weak--emission and classical T Tauri stars in Taurus, 
Skrutskie {\it et al.} (1990) estimated the timescale for
dissipation of accretion disks around young stars to be 
$< 10^7$ yrs.  They also identified three stars as ``transition
objects'', thought to be in the process of dissipating an 
optically--thick disk.  Based on the small number of these
objects, Skrutskie {\it et al.} derive a timescale
of $< 1 \times 10^6$ yr for transition from optically--thick
accretion disk to optically--thin re-processing disk 
(see also Wolk and Walter, 1996). 
These timescales are important for constraining the 
epoch of planet formation and providing
insight into the disk dissipation process.  For example, 
calculations by Pollack {\it et al.} (1997) require mass
surface densities 3--4 times greater than implied
by the minimum mass solar nebula for Jupiter to form  in $< 10$ Myr 
via runaway accretion. Because of the
small sample ($\sim 20$ objects) observed from the ground 
at 10 $\mu$m, as well as the sensitivity limitations of 
the IRAS satellite at 12 $\mu$m, Skrutskie {\it et al.} 
were unable to constrain the lifetime of optically--thin 
circumstellar disks in the terrestrial planet zone.

We have begun a program to determine the frequency of 
optically--thin 10 $\mu$m emission among T Tauri stars in the Taurus 
dark cloud utilizing the current generation of mid--IR detectors with 
sensitivity limits $\times 10$ better than that of the IRAS satellite.  
We hope to learn whether or
not the termination of active disk accretion is also accompanied by
rapid clearing of the inner--disk.  Here we describe initial results  
obtained for one of the transition objects identified by Skrutskie
{\it et al.} (1990), DI Tau, which is spatially unresolved from the nearby
classical T Tauri star DH Tau in the IRAS beam.  We present new 
ground--based 10 $\mu$m observations resolving the emission from both
stars, construct updated spectral energy distributions (SEDs) 
for both objects, and re--analyze their stellar and circumstellar disk 
properties.  Our analysis shows that DI Tau does not possess
an optically--thick circumstellar disk within 0.1 A.U. of the central 
star and that its previously known companion, 
located at a distance of $\sim$ 20 A.U., could very well be a brown dwarf.
We speculate that the formation of a very low mass companion orbiting
DI Tauri led to the rapid evolution of its inferred circumstellar disk. 

\section{New Mid--Infrared Observations} 

The data were obtained with
the Mid--infrared Array eXpandable (MAX) camera
constructed by Infrared Labs for the Max--Planck--Institut f\"ur 
Astronomie. The MAX camera is built around a Rockwell 128$\times$128
Si:As BIB array which provides a field of view 35'' $\times$
35'' when mounted on the 3.8m United Kingdom Infrared Telescope (UKIRT).
Observations were made at UKIRT on August 26--27, 1996
during photometric conditions with diffraction--limited images (FWHM $\sim 0.7$'') 
obtained through an N--band filter ($\lambda_{eff} = 10.16 \mu$m; 
$\Delta \lambda = 5.20 \mu m$). Data were collected while
chopping the telescope N--S (12'') at a rate of 2 Hz, and nodding
the telescope (12'') every 50 seconds to correct for non--uniform 
illumination effects introduced by chopping.  Data were reduced according to 
standard image processing techniques except that no flat--field 
corrections were applied. Images obtained at each end of the ``chop''
were subtracted from each other to remove bias, 
dark current, and thermal
background.  Co--added images from both ``nod'' positions were averaged 
and aperture photometry was performed on the final images with a 
diameter of 3.12'' using a sky annulus of 5.2--10.4''.  
Flux calibration was derived 
by observing standards from the list of Cohen {\it et al.}
(1992).  Both DI and the nearby DH Tau ($sep = 15.1$''; $PA = 307^{\circ}$) were
observed simultaneously on the array ($T_{int} = 250.0$s), 
interspersed with observations of the standard star HR1370 ($T_{int} = 50.0$s) 
at nearly the same airmass ($\Delta X < 0.1$).
Comparison of photometry from stellar images
appearing on different portions of the array indicates residual uncertainties 
in the calibration less than $\pm 5$\%. Derived fluxes and associated errors 
(dominated by the thermal background) are:
DH Tau $F_N = 0.137 \pm 0.005$ Jy (6.26$^m$) and DI Tau $F_N = 0.030 \pm 
0.005$ Jy (7.90$^m$). Additional observations were obtained in the 
Q--band ($\lambda_c =  19.91 \mu$m; $\Delta \lambda = 1.88 \mu m$).
during non--photometric conditions yielding a flux ratio of $> 1.7$ 
between DH Tau (detected) and DI Tau (undetected). 

\section{Revised Spectral Energy Distributions and Stellar Parameters}

We combine the new photometry described above with previously
published simultaneous optical and infrared data compiled by 
Rydgren and Vrba (1981) from 0.3--3.8 $\mu$m to
construct updated SEDs for these 
sources.  We adopt the spectral types listed in
Cohen and Kuhi (1979) as well as the IRAS fluxes 
recently derived by Beckwith {\it et al.} (1997).  Because the 
mid--IR flux of DH Tau dominates that of DI Tau by factors of 
four and two at 10 and 20 $\mu$m respectively, 
we associate the IRAS flux with DH Tau.  In order to 
de-redden the observed spectral energy distribution, we use the color
excess observed in the $(R-I)_c$ index and adopt the
reddening law of Rieke and Lebofsky (1985), transformed into the 
appropriate color system.  The stellar contribution is estimated 
by normalizing the dereddened I--band flux to that expected from 
a dwarf star of the same spectral type. Key stellar and circumstellar parameters are
summarized in Table~\ref{props} for both objects. 

\begin{figure}
\insertplot{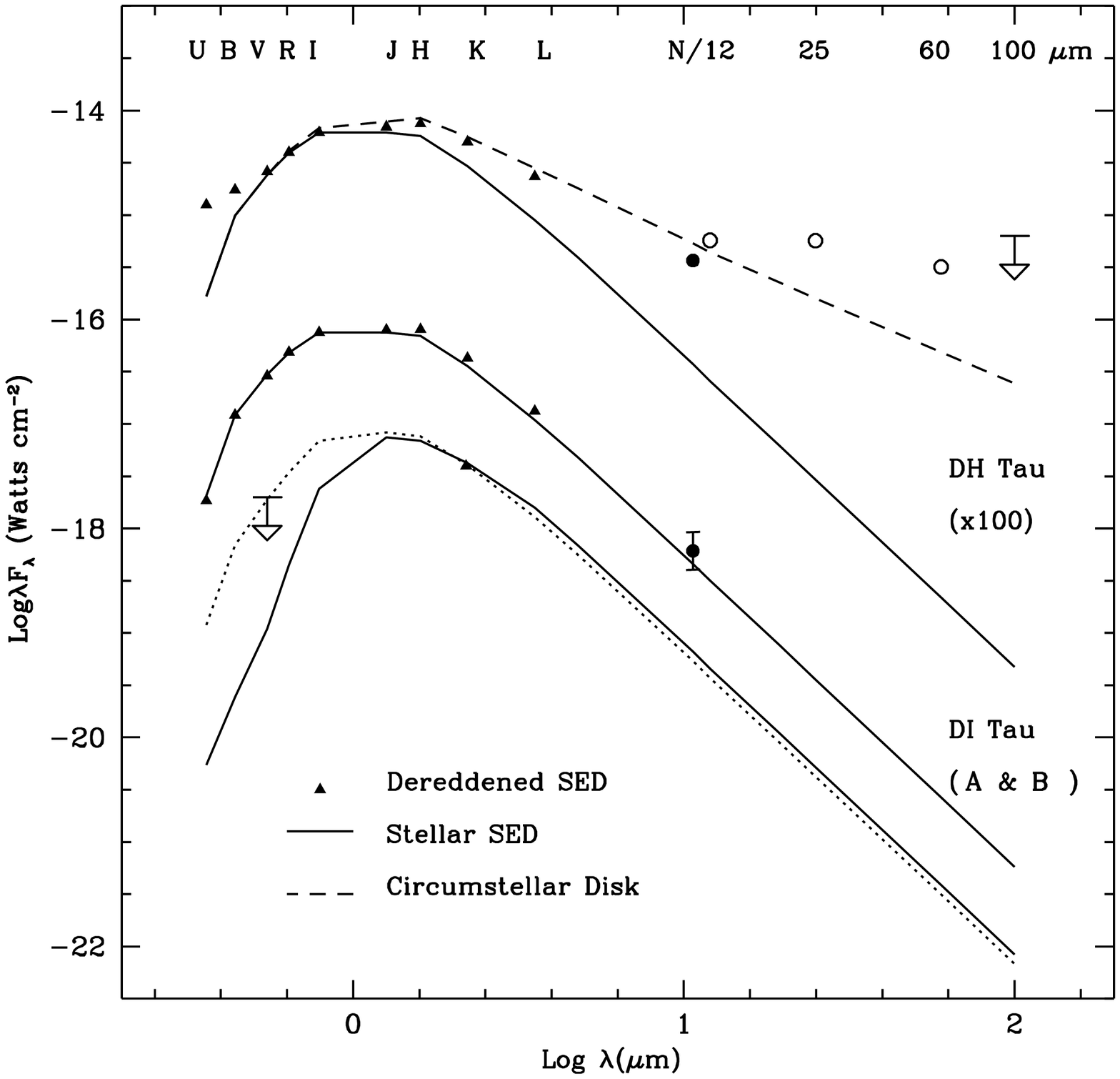}{8.0}{10.}{-0.75}{1.0}{0.7}{0}
\caption[]{ Spectral energy distributions for both DH and
DI Tauri:  The filled--triangles represent simultaneous optical/near--IR
data taken from Rydgren and Vrba (1979) and de-reddened as described
in the text, the filled--circles represent
our new 10 $\mu$m observations, and the open--circles are IRAS fluxes from
Beckwith {\it et al.} (1997).  Arrows indicate upper--limits.
Unless otherwise indicated, the observational errors are
smaller than the points.  The solid lines are the
expected photospheric emission; the dashed--line is the emission expected
from a face--on re-processing disk which extends into the stellar surface.
While the SED for DH Tau exhibits near-- and mid--IR excess emission
consistent with an optically--thick circumstellar accretion disk, the circumstellar
environment of DI Tau A appears to be free of material within 0.1 AU (10 R$_*$).
The SEDs for DI Tau B are shown normalized at 2.2 $\mu$m for
an intrinsic spectral type of M2 (dotted--line corresponding to an absolute 
upper limit to the V--band flux)
and M6 (solid line assuming the primary and secondary are co--eval).
}
\label{seds}
\end{figure}

The de-reddened SEDs are shown in Figure~\ref{seds} along with those expected 
from stellar photospheric emission 
and a face--on re-processing disk model (e.g. Hillenbrand {\it et al.}, 
1992).  As mentioned above, DI Tau does not exhibit significant 
infrared excess emission out to a wavelength of 10 $\mu$m 
while DH Tau shows both ultraviolet and infrared excess
emission typical of classical T Tauri stars thought to possess 
active accretion disks.  This is consistent with recently published 
spectroscopic studies of both stars:
Hartigan, Edwards, and Ghandor (1995) place an upper--limit on the mass 
accretion rate of DI Tau at $< 1.5 \times 10^{-8} M_{\odot} yr^{-1}$
while Valenti, Basri, and Johns (1993) detect significant accretion 
luminosity in DH Tau. Comparison of the SED with blackbody models of 
optically--thick circumstellar disk emission suggests that if 
DI Tau does possess a disk, it must be evacuated within at 
least $10 R_*$ ($\sim$ 0.1 A.U.).
DH Tau appears to have a disk which extends to within a few
stellar radii (depending on the inclination angle and disk 
accretion rate adopted; e.g. Meyer {\it et al.}, 1997). 

\begin{figure}
\insertplot{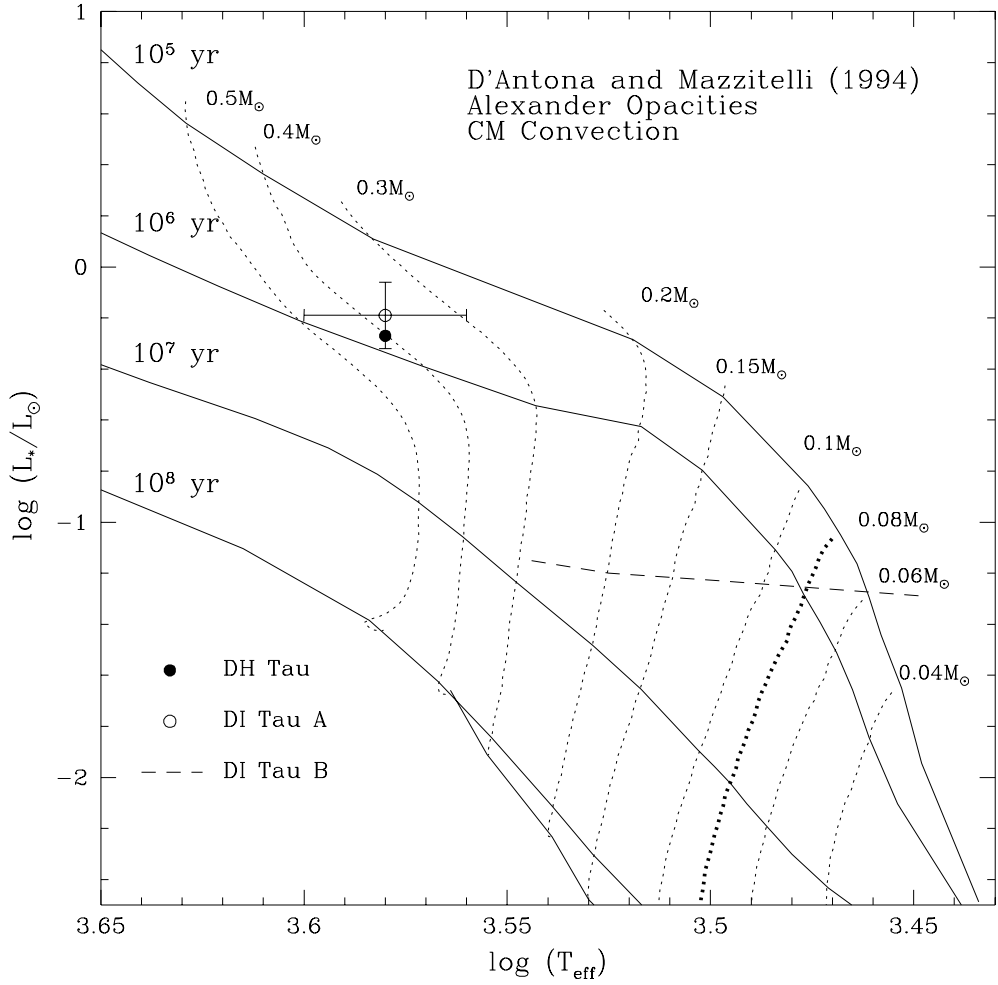}{8.0}{10.}{-0.75}{1.0}{0.7}{0}
\caption[]{The H--R Diagram for DH Tau and DI Tau A along with the PMS evolutionary
models of DM94.  The track corresponding to the hydrogen--burning limit is
indicated at 0.08 M$_{\odot}$.  Also shown is the range of effective temperatures
and luminosities derived for DI Tau B.  The error-bars for the point
corresponding to DI Tau A indicate an age $< 10^6$ yrs.  If the companion
DI Tau B is co--eval according to these isochrones, then it must
have a spectral type $>$ M5 and a mass $< 0.08 M_{\odot}$.
}
\label{hrd}
\end{figure}

Using the effective temperatures and luminosities listed
in Table~\ref{props}, we place the stars in the H--R diagram
(Figure ~\ref{hrd}) 
for comparison with the PMS evolutionary models of D'Antona and Mazzitelli
(1994; hereafter DM94) adopting the Alexander opacities and Canuto--Mazzitelli
convection prescription.  
DH Tau and DI Tau are very young ($< 10^6$ yrs), low--mass 
($< 1.0 M_{\odot}$) PMS objects. 
From examining the properties listed in Table~\ref{props}, 
it is clear that DH and DI Tau are quite similar 
except in two fundamental respects: 
DH Tau is a single star with an active accretion disk
and DI Tau is a binary system lacking such a disk. 

\section{A Sub-stellar Companion to DI Tauri?}

Both stars were part of the lunar occultation, speckle, and 
direct imaging survey of Simon {\it et al.} (1995) to measure 
the binary frequency of pre--main sequence systems. 
DH Tau has no companions between 0.005-10'' 
(1--1400 A.U. assuming a distance of 140 pc to Taurus). 
DI Tau A has a companion (hereafter referred to as B) 
at a projected separation of 
0.12'' (16.8 A.U.) with a K--band flux 
ratio of $8 \pm 1$ (Ghez {\it et al.}, 1993).
The probability of observing a chance projection of a field 
star with $K < 12.0^m$
at this separation is $< 2 \times 10^{-6}$.  
Recent HST observations of DI Tau by Simon {\it et al.} (1996) 
with the Fine Guidance Sensor 
provide a lower limit to the V--band flux ratio of the 
system; $\Delta V > 3.3^m \pm 0.3$
\footnote{Although not discussed explicitly in Simon {\it et al.}
(1996), the simulations presented in Lattanzi {\it et al.} (1992) 
provide an estimate of the error associated with this magnitude
difference given the separation of the system (see their Figure 3).}. 
Because of the systematic uncertainties in the FGS photometry reported
in Simon {\it et al.} (1996), we adopted the mean magnitude (and range) 
derived for DI Tauri from over a decade of photometric monitoring 
(e.g. Herbst {\it et al.}, 1994) of 
$<m_V> = 12.85$ ($\delta m_v = 0.14^m$) resulting in $m_V > 16.2^m \pm 0.3$ for
the companion.  
Adopting the extinction derived for the primary (with uncertainty of $\pm 0.5^m$) 
in the DI Tau system, the {\it lower limit} to the intrinsic color
of the companion is $(V-K)_o > 4.6^m \pm 0.5$ indicating that the 
companion must have a spectral type of M2 or later.  Given this 
constraint on the effective temperature, the range of associated
luminosities for the companion is shown as the dotted--line 
in Figure ~\ref{hrd}.  From the position of DI Tau A in the H--R 
diagram, its age must be $< 1 \times 10^6$ yr.  
If the system is co--eval according to the DM94 
tracks, the upper--limit on the age of the primary 
implies a companion mass of $< 0.08 M_{\odot}$
(corresponding to a spectral type of M5)
\footnote{The uncertainty in the luminosity estimate of DI Tau B is much 
smaller than that for DI Tau A where we associated the errors in extinction 
and distance with the primary.}.  
Adopting the models of Burrows {\it et al.} (1997), an
age $< 10^6$ yrs, and stellar luminosity $< 0.06 L_{\odot}$ implies a
companion mass $< 0.08 M_{\odot}$.  How likely is it that 
binary is co--eval?  Hartigan, Strom, and Strom (1994) 
find that $2/3$ of the PMS binaries in their sample (separations between 400--6000 AU)
are co--eval when compared to the evolutionary models of DM94.  
In cases where an age difference is observed, the lower mass companion 
always appears younger.  Brandner and Zinnecker (1997) find that
all binaries in their sample (90--250 AU) are co--eval within the observational errors. 
The lack of infrared excess observed out to 10 $\mu$m for 
the DI Tau system precludes the possibility that the low luminosity 
companion detected at 2.2 $\mu$m is in a different 
evolutionary state than the primary (i.e. is an infrared 
companion).  The spectral index ($\lambda F_{\lambda} \sim \lambda^{-\alpha}$) 
of the companion must be $\alpha > 4/3$ from 2--10 $\mu$m. 
The luminosity ratio between the two stellar components is 
$> 10$, one of the highest known among the very young low mass 
stars in the Taurus dark cloud.  Given that brown dwarf companions have 
been discovered around low mass stars in the solar neighborhood 
(GL229B, Nakajima {\it et al.}, 1995; HD114762, Latham {\it et al.}, 1989), 
we consider it reasonable to postulate their existence in the 
pre--main sequence; the companion to DI Tau A could very well be a brown dwarf 
\footnote{Ghez {\it et al.} (1997) has recently studied several
T Tauri binary systems and uncovered three additional candidate
brown dwarf companions.}.  In fact DI Tau bears a striking resemblance 
to what the GL229 system might have looked like at an age of 
$< 10^6$ yrs.  This hypothesis could be tested by
confirming that DI Tau B has a spectral type later than M5, or measuring orbital 
motions through monitoring of the relative positions of the DI Tau 
system (e.g. Ghez {\it et al.}, 1995).  


\section{Evidence for Rapid Disk Evolution for DI Tau A} 

In the preceding discussion, we have demonstrated that DI Tau A
does not possess an infrared excess indicative of an optically--thick 
inner circumstellar disk and that the previously known companion could be 
a brown dwarf orbiting at a distance of $\sim$ 20 A.U. from the central star.  
 Given the inferred age  
($\sim 6 \times 10^5$ yrs), if there was a disk present
within 0.1 AU around this star in the past, its lifetime was very short.
Arising from the collapse of a rotating cloud core, disks 
are expected to serve as the main reservoir of angular momentum in 
young stellar objects, as Jupiter does in our own solar system.  
Perhaps DI Tau A never had a substantial circumstellar disk, 
the excess angular momentum being stored in the orbit of the system. 
Indeed, the DI Tau binary harbors $\sim 100$ times the angular momentum
of the DH Tau star+disk system even though, separated by only 2100 AU, 
they presumably formed from the same parent molecular cloud core. 
Yet DI Tau A is the most slowly rotating star known in the Taurus
dark cloud that does not exhibit 10 $\mu$m excess emission
(Meyer and Beckwith, 1997). 
Can this tell us something about the history of the circumstellar environment? 
Edwards {\it et al.} (1993; see also Bouvier {\it et al.}, 1993)
have presented evidence that stellar angular momentum is regulated
by the presence of a circumstellar disk.  This  
explains the slow rotation rates of classical T Tauri stars ($P > 5$ days) 
compared to the weak--lined T Tauri stars which rotate faster 
($P < 5$ days).  Their discovery finds theoretical support in models of
magnetospheric coupling between young stars and circumstellar 
accretion disks (K\"onigl, 1991; Shu {\it et al.} 1994; Cameron, {\it et al.}, 1995). 
The Kelvin--Helmholtz timescale for a disk-less star to spin--up from
8 days to $<$ 6 days is is roughly $1 \times 10^5$ yrs  (Armitage
and Clark, 1996).  Tidal effects in the DI Tau system are negligible since 
the the semi--major axis is very large compared to the radii of the objects 
(e.g. Rasio {\it et al.}, 1996). 
The disk--regulated angular momentum hypothesis implies that, 
given the rotation rate of DI Tau A ($P = 7.9 \pm 0.5$ days),
this star probably had a circumstellar disk in the very recent past. 

Could the formation of a very low mass companion have contributed to 
the dissipation of a circumstellar disk surrounding DI Tau A?
Artymowicz and Lubow (1994) have suggested that gaps in circumstellar 
disks can be created due to dynamical clearing by a binary companion. 
The typical distances spanned by these gaps are from half to twice the
semi--major axis (10--40 A.U. for the DI Tau system). 
An inner accretion disk not fed by an outer disk would dissipate very quickly. 
For a typical disk mass of 0.02 M$_{\odot}$ (Osterloh and Beckwith, 1995)
the amount of material located within 10 A.U. is $\sim 0.005 M_{\odot}$.
The lifetime of such an inner--disk given a typical disk accretion rate of 
$10^{-7}$ M$_{\odot} yr^{-1}$ (Hartigan, Edwards, and Ghandour; 1995)
would be $<$ 50,000 yrs.  A cold outer disk 
might still be present albeit of very low mass ($M_D < 0.001 M_{\odot}$; 
Jensen {\it et al.}, 1994).   Is it possible that DI Tau A remains locked
to a remnant disk located between the inner edge derived here ($> 0.1 AU$) 
and the tidal truncation radius of $\sim 10 AU$ ?  Armitage and Clarke (1996) 
have suggested this might occur in binary systems with separations between 1--8 AU, 
where material is trapped between the co-rotation point of the star--disk 
system ($6R_*$ for DI Tau A) and the inner 
tidal truncation radius ($0.5 \times sep = 10.0 AU$).
This material serves to transfer 
angular momentum from the central star as it contracts to the orbit of the 
binary companion, resulting in slower rotation rates for binary stars 
separated by a few AU compared to single stars or wide binaries. 
Future observations at wavelengths $> 50 \mu$m and spatial resolution $< 15$''
(e.g. with SOFIA) will be required to set upper limits $< 10^{-4} M_{\odot}$ 
on any remnant material orbiting the DI Tau system.

More recent work by Artymowicz and Lubow (1996) suggests that
material from a circumbinary disk could still move across the tidal gap.
This material would preferentially accrete onto the lower mass component of the
system, tending to drive the mass ratio towards unity. 
This suggests that if the sub--stellar mass companion
formed in the disk of DI Tau A, it did so after the main infall phase
had ended, with $< 0.1 M_{\odot}$ remaining in the outer disk. 
Run--away giant planet growth in a proto--planetary 
disk (e.g. Pollack {\it et al.}, 1997) can probably be ruled out
given the formation time of $< 10^6$ yrs.  Perhaps the
companion to DI Tau A formed rapidly through gravitational instabilities in a massive 
circumstellar disk (e.g. Boss, 1997).  A more conventional 
explanation would be that the DI Tau system simply formed from the 
fragmentation of a rotating collapsing cloud core 
(e.g. Burkert and Bodenheimer, 1997).  If DI Tau A had a circumstellar
disk, it seems reasonable to conclude that 
the presence of sub--stellar mass companion at 20 A.U.  
contributed to its rapid evolution. 

Ghez {\it et al.} (1994; see also Jensen {\it et al.}, 1996) 
have offered a similar explanation for the 
differences observed between UZ Tau East 
(a single star with optically--thick disk recently discovered to be a 
spectroscopic binary!) 
and UZ Tau West (a binary system with optically--thin disk). 
Is there a connection between binarity and the lifetime of circumstellar 
disks?  Osterloh and Beckwith (1995; see also Jensen, Mathieu, and Fuller, 
1994) find that young binaries with separations
$< 100$ A.U. emit less at mm--wavelengths than young single stars 
and wide binaries. However, Simon and Prato (1995) find no correlation 
between the presence of companions between 20--280 A.U. and 
the presence of an inner--disk.  This finding is confirmed by examining the
distribution of $(K-N)$ excesses vs. binary separation (Meyer and Beckwith, 1997)
including newly discovered CTTS systems with separations $< 1.0$ AU 
(such as UZ Tau East and DQ Tau; Mathieu {\it et al.}, 1997). 
Although there is circumstantial evidence in at 
least at least two cases that the presence of a very low mass companion may have
influenced the evolution of a circumstellar disk, it remains
an open question over what range of separations (and mass ratios) 
the influence of a binary companion is important. 

\section{Acknowledgements}
 
We would like to thank Peter Bizenberger, Christoph Birk, and the staff of 
UKIRT for their help in commissioning 
the MAX camera.  We also thank the referee, M.F. Skrutskie, for a number of 
insightful comments which improved the manuscript. 
We gratefully acknowledge fruitful discussions with Mathew Bate, Suzan Edwards, 
Lynne Hillenbrand, Christoph Leinert, and Geoff Marcy.  Special thanks to
LAH for assistance with the sub-stellar evolutionary tracks.

\newpage

\newpage

\begin{table}
\begin{small}
\centering
\caption{Stellar and Circumstellar Properties}
\begin{tabular}{lllllllllll}
\multicolumn{1}{c}{Name}   &
\multicolumn{1}{c}{T$_* K^1$} &
\multicolumn{1}{c}{A$_V$}  &
\multicolumn{1}{c}{$L_*/L_{\odot}$} &
\multicolumn{1}{c}{$M_*/M_{\odot}$} &
\multicolumn{1}{c}{$\tau$ (yrs)} &
\multicolumn{1}{c}{Sep.$^2$} &
\multicolumn{1}{c}{Per. $^3$}   &
\multicolumn{1}{c}{$(K-L)_o^4$}   &
\multicolumn{1}{c}{$\Delta N^5$}   &
\multicolumn{1}{c}{$M_D/M_{\odot}^6$ }  \\\hline\hline
DH Tau & 3800 & 1.7$^m$ & 0.54  & 0.40 & 8$\times 10^5$ & NC & 7.2d & $0.60^m$ & $0.94 \pm 0.16$ & 0.011 \\
DI Tau & 3800 & 0.8$^m$ & 0.65  & 0.38 & 6$\times 10^5$  & 0.12'' & 7.9d & $0.16^m$ & $0.16 \pm 0.18$ & $<$ 0.001 \\\hline
\multicolumn{11}{l}{\footnotesize $^1$
Data taken from Cohen and Kuhi (1979) for DH Tau (HBC \# 38) and DI Tau (HBC \# 39). } \\
\multicolumn{11}{l}{\footnotesize $^2$
Presence or absence of companions taken from Simon et al. (1995). } \\
\multicolumn{11}{l}{\footnotesize $^3$ Photometric rotation periods taken from Vrba {\it et al.} (1989). } \\
\multicolumn{11}{l}{\footnotesize $^4$
De-reddened $(K-L)$ color using A$_V$ listed here; errors $\pm 0.11^m$.} \\
\multicolumn{11}{l}{\footnotesize $^5$
N--band excess in dex as defined by Skrutskie {\it et al.} (1990). } \\
\multicolumn{11}{l}{\footnotesize $^6$
Disk masses taken from Dutrey {\it et al.} (1996) for DH Tau and Jensen {\it et al.} (1994)
for DI Tau.} \\
\label{props}
\end{tabular}
\end{small}
\end{table}

\end{document}